\documentclass{aa}  

\usepackage{amsmath} 
\usepackage{amssymb} 
\usepackage{xcolor}
\usepackage[]{geometry}
\usepackage{txfonts}
\usepackage[]{hyperref}
\usepackage{verbatim}
\usepackage{enumitem}
\usepackage{epsfig}
\usepackage{multirow}
\usepackage{tablefootnote}
\usepackage{stfloats}

\usepackage{graphicx,subcaption}

\def\erosita{eROSITA}
\def\rosat{ROSAT}
\def\os{{O~{\sc vii}}}
\def\oe{{O~{\sc viii}}}

\newcommand{\degree}{\ensuremath{^\circ}}

\def\be{\begin{equation}}
\def\ee{\end{equation}}

\begin{document}
\titlerunning{Soft X-ray emission lines}
\authorrunning{Zheng et al.}

\title{eROSITA narrowband maps at the energies of soft X-ray emission lines}
\subtitle{}
\author{
Xueying Zheng \inst{1}, 
Gabriele Ponti \inst{2,1},
Nicola Locatelli \inst{2,1}\thanks{emails to: nicola.locatelli@inaf.it},
Jeremy Sanders \inst{1},
Andrea Merloni \inst{1},
Werner Becker \inst{1},
Johan Comparat\inst{1},
Konrad Dennerl \inst{1},
Michael Freyberg \inst{1},
Chandreyee Maitra\inst{1},
Manami Sasaki \inst{3},
Andrew Strong\inst{1}
Michael~C.~H.~Yeung \inst{1}
}

\institute{
     Max-Planck-Institut f{\"u}r extraterrestrische Physik, Gießenbachstraße 1, D-85748 Garching bei M\"unchen, Germany
\and INAF-Osservatorio Astronomico di Brera, Via E. Bianchi 46, I-23807 Merate (LC), Italy
\and Dr. Karl Remeis Observatory, Erlangen Centre for Astroparticle Physics, Friedrich-Alexander-Universit\"at Erlangen-N\"urnberg Sternwartstraße 7, 96049 Bamberg, Germany
}

    \date{}
    \abstract{
Hot plasma plays a crucial role in regulating the baryon cycle within the Milky Way, flowing from the energetic sources in the Galactic center and disc, to the corona and the halo. This hot plasma represents an important fraction of the Galactic baryons, plays a key role in galactic outflows and is an important ingredient in galaxy evolution models.
Taking advantage of the \textit{Spectrum-Roentgen-Gamma (SRG})/\erosita~first all-sky survey (eRASS1), in this work, we aim to provide a panoramic view of the hot circumgalactic medium (CGM) of the Milky Way. 
Compared to the previous all sky X-ray survey performed by ROSAT, the improved energy resolution of \erosita\ enables us to map, for the first time, the sky within the narrow energy bands characteristic of soft X-ray emission lines.
These provide essential information on the physical properties of the hot plasma. 
Here we present the \erosita\ eRASS1\ half sky maps in narrow energy bands corresponding to the most prominent soft X-ray lines: \os\ and \oe, which allow us to constrain the distribution of the hot plasma within and surrounding the Milky Way. 
We corrected the maps by removing the expected contribution associated with the cosmic X-ray background, the time-variable solar wind charge exchange, and the local hot bubble. We applied corrections to mitigate the effect of absorption, therefore highlighting the emission from the CGM of the Milky Way. 
We use the line ratio of the oxygen lines as a proxy to constrain the temperature of the warm-hot CGM, and we define a pseudo-temperature $\mathcal{T}$ map. The map highlights how different regions are dominated by different thermal components.
Towards the outer halo, the temperature distribution of the CGM on angular scales of 2-20 deg is consistent with being constant $\Delta \mathcal{T} / \langle \mathcal{T}\rangle \leq 4\%$, with a marginal detection of $\Delta \mathcal{T} / \langle \mathcal{T}\rangle = 2.7 \% \pm 0.2\%$ (statistical) $\pm 0.6\%$ (systematic) in the southern hemisphere. Instead, significant variations $\sim 12\%$ are observed on many tens of degrees scales when comparing the northern and southern hemispheres.
The pseudo-temperature map shows significant variations across the borders of the \erosita\ bubbles, suggesting temperature variations, possibly linked to shocks, between the interior of the Galactic outflow and the unperturbed CGM. 
In particular, a "shell" characterized by a lower line ratio appears close to the edge of the \erosita\ bubbles.
  }
  \keywords{diffuse radiation—Galaxy: center—surveys—X-rays: galaxies—X-rays: ISM}
    \maketitle

\section{Introduction}

As the next generation of all-sky survey experiments in soft X-ray, the extended ROentgen Survey with an Imaging Telescope Array (\erosita), aboard the Russian-German observatory {\it Spectrum-Roentgen-Gamma} \citep[{\it SRG};][]{Sunyaev2021}, has performed the first all sky survey (eRASS1; \citealt{Merloni24}) from December 2019 until June 2020. eRASS1 has an unprecedented sensitivity to the whole sky in the soft X-ray range, thanks to an effective area of $\approx$1100 cm${}^2$ at 1.4 keV, with an energy resolution typical of CCD instruments \citep[$\approx$ 65 eV at 0.5 keV;][]{Predehl2021}.
Hence \erosita\ makes it possible to obtain "monochromatic" images for various emission lines simultaneously from the diffuse emission over the entire sky.
 
In this paper, we present the eRASS1 narrowband maps corresponding to the brightest transitions in the soft X-ray band (i.e., \os\ and \oe), together with the maps ratio, and make these maps (in count rate unit), publicly available online\footnote{\url{https://erosita.mpe.mpg.de/dr1/AllSkySurveyData_dr1/HalfSkyMaps_dr1/}}.

In X-rays, emission lines are vital for understanding the properties of the hot plasma inside the interstellar medium (ISM) and in the circum-Galactic medium (CGM), the latter made up of different components, among which the Galactic halo, the Galactic corona and the Galactic outflow are also included. 
The halo of the Milky Way is expected to be filled with plasma with temperatures close to its virial temperature\footnote{For simplicity, we refer to $kT$ as temperature or just $T$ throughout the paper. Therefore, we report all temperatures $T$ using keV units.}, $T\sim0.15-0.20$ keV, therefore appearing bright in soft X-ray emission lines \citep{Yoshino2009PASJ, Henley2010ApJ, Henley2012ApJS, Miller2015ApJ}.
Additionally, Galactic outflows emerging from the center of the Milky Way are expected (and observed) to drive shocks into the ISM and the unperturbed CGM, heating the plasma and inducing bright soft X-ray emission lines \citep{Predehl2020Natur}.
Finally, it has been argued that activity in the Galactic disc is expected to sustain a hot Galactic atmosphere (the so called `Galactic corona'), which might be also traced in the soft X-ray band \citep{2019ApJ...882L..23D, 2019ApJ...887..257D, Ponti2023}.

Mapping the emission lines from such hot and rarefied plasma is extremely challenging because the corresponding signal is spread over the entire sky and at very low surface brightness.
To increase sensitivity to this emission, broad band images can be helpful.
The ROSAT\ all sky survey allowed us to get a view of the CGM emission in broad energy bands \citep{Snowden1997ApJ, Zheng2023}, however the \os\ and \oe\ lines (i.e. the brightest) could not be imaged separately because of the limited energy resolution of ROSAT.
With its high sensitivity and band coverage in soft X-ray, \erosita\ possesses the specialized capability to image the faint emission in narrower energy bands compared to ROSAT{}.
Here we present the first \erosita\ sky maps of the soft X-ray emission lines.

The paper is organised as follows:
Section \ref{sec:data} shows the data analysis; Section \ref{sec:lines} discusses the soft X-ray emission lines as tracers of hot plasma; Section \ref{sec:subtract} presents the necessary tools that have been used to retrieve physical information from the maps;
Section \ref{sec:narrowband_maps} discusses the morphology of the line emission maps, their ratio, and the retrieved physical information. Section~\ref{sec:ratios} discusses the observed ratio between emission line maps as a proxy for the plasma temperature. Section~\ref{sec:CGM_T} and \ref{sec:bubbles_T} present the constraints on the temperature of the CGM and the \erosita\ bubbles, respectively. We further discuss our findings in Sect.~\ref{sec:discussion}, and finally summarized in Section \ref{sec:sum}.

\section{Data} \label{sec:data}

We analysed the eRASS1 data \citep{Merloni24}, collected from December 12, 2019 until June 11, 2020. 
We considered only data from the five telescope modules (TM) of \erosita\ cameras equipped with on-chip filter (i.e., TM1, TM2, TM3, TM4, TM6), to avoid any contamination from the effects of light leak (affecting TM 5 and 7, \citealt{Predehl2021}).
We used the same procedures as in \cite{Zheng2023} and analyze the data with the standard \erosita\ Science Analysis Software System (\texttt{eSASS}), version 020, developed by the German \erosita\ consortium \citep{2022A&A...661A...1B}.
The \erosita\ data of the all sky are archived in a system of 4700 partially overlapping tiles of $3\fdg6\times 3\fdg6$ each \citep{Merloni24}. 
When required, we re-projected each pixel in each sky-tile, to create a map with an HEALPix\footnote{\url{ http://healpix.sourceforge.net}} projection with \texttt{NSIDE}=512 \citep{Gorski2005ApJ, Zonca2019}.
The same filtering described in Zheng et al. (2023) has been applied to the counts maps and exposure maps. 

The maps display the intensity within the specific energy range in unit of counts per second per square degree $[\rm cts\, s^{-1}\, deg^{-2}]$. 
The Zenital Equal Area (ZEA) projection is applied to conserve surface brightness. While the ZEA projection introduces projection distortion, especially towards the borders of the map, we find the distortions less prominent compared to other kinds of possible projections. For display purposes only, several maps presented below were smoothed with an adaptive kernel based on a count rate S/N threshold of 15 to highlight the diffuse emission.

\section{Soft X-ray lines as tracers of hot plasma}
\label{sec:lines}

\begin{figure}
    \centering
    \includegraphics[width=0.5\textwidth,angle=0]{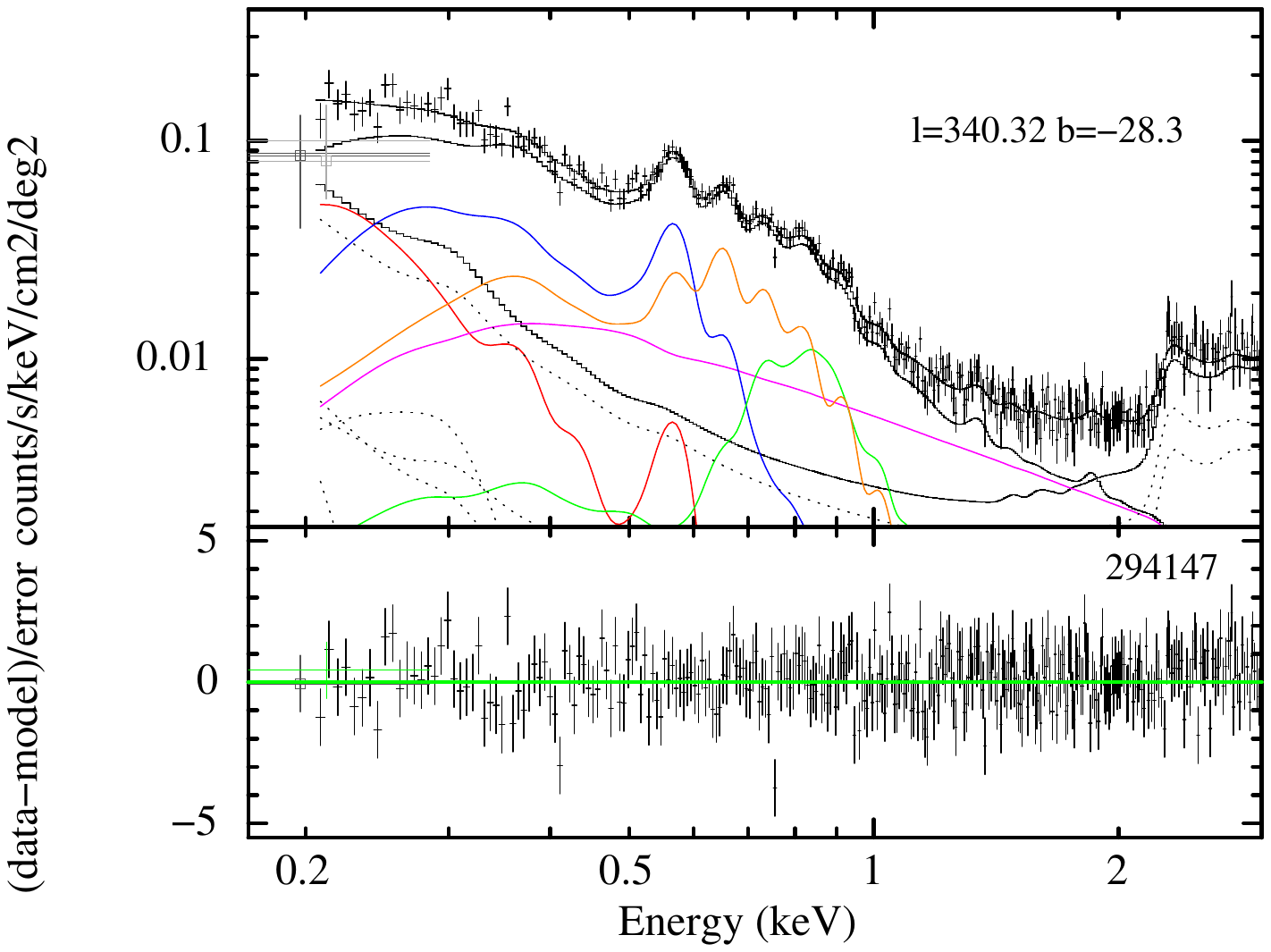}
    \includegraphics[width=0.5\textwidth,angle=0]{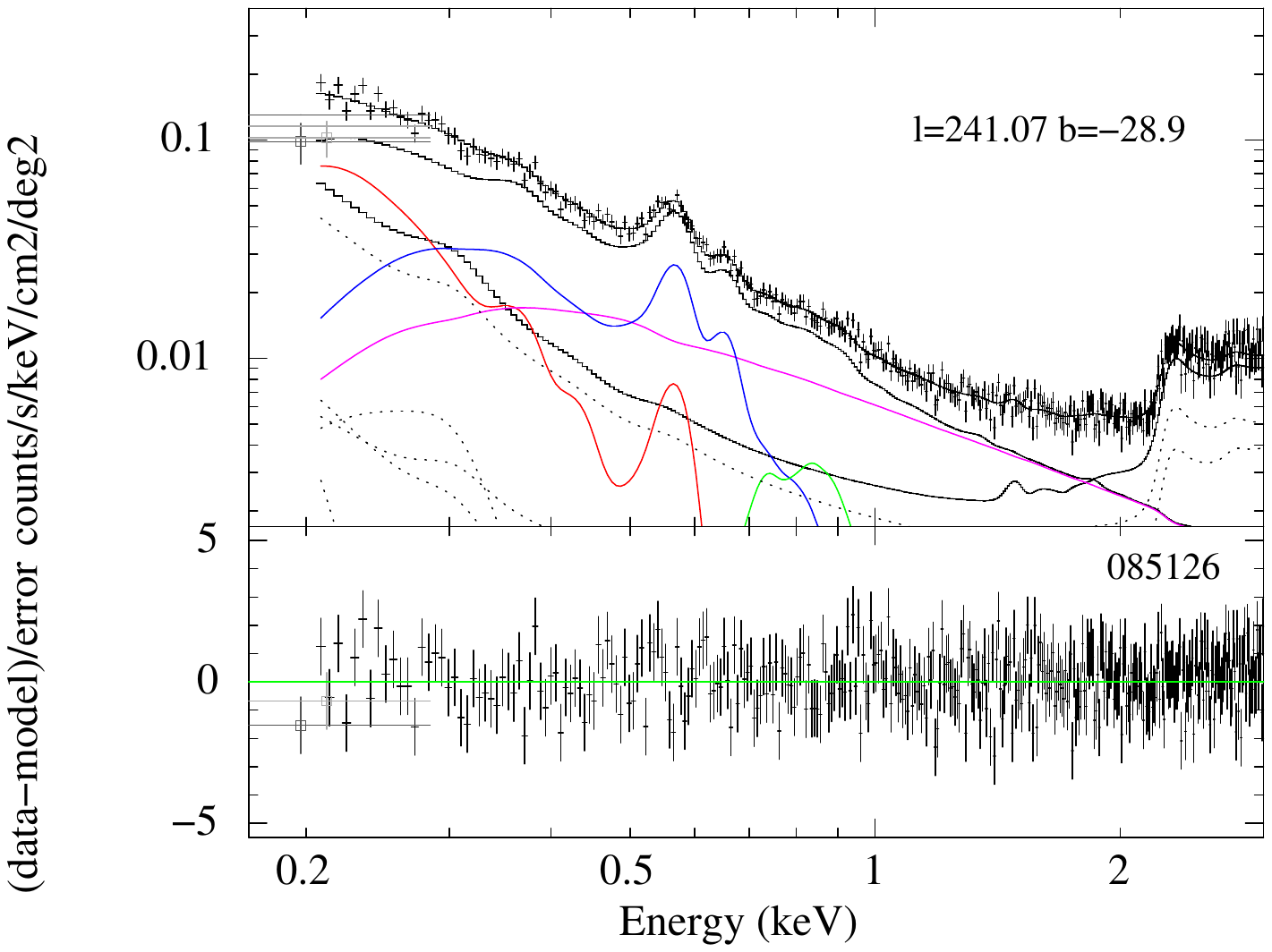}
    \caption{Spectra of the diffuse emission observed by \erosita\ within two sky patches of $3\degree \times 3\degree$ each.
    Upper panel: observed and modeled spectrum of the sky region at $l=340\fdg32$, $b=-28\fdg3$ (eROSITA sky tiles 294147). 
    Lower panel: observed and modeled spectrum of the sky region at $l=241\fdg07$, $b=-25\fdg9$ (eROSITA sky tiles 085126).
    The regions have been chosen to show the typical diffuse emission from a patches of the sky, with(out) the contribution from the \erosita\ bubbles emission. The contribution from the detected point sources has been removed. The grey dots at $E=0.2$ keV show the ROSAT R1 and R2 data points. The black solid and dotted lines show the contribution of the instrumental background (derived by the fit of the filter wheel closed data). The red, blue, green and magenta lines show the contribution of the LHB, warm-hot CGM, hot corona and CXB emission components. The orange line (upper panel only) shows an additional plasma component that can be introduced in the \erosita\ bubble region.}
    \label{fig:spec}
\end{figure}

The electromagnetic spectrum of optically thin and collisionally ionised (hot) plasma consists of continuum emission plus an array of emission lines.
For X-ray emitting plasma with temperatures lower than $T\sim1$~keV, the majority of the flux is in fact carried by a large number of weak emission lines. 
The numerous lines cannot be resolved individually at the CCD energy resolution of \erosita, forming a pseudo continuum. 
As an example, the data in the top and bottom panels of Fig. \ref{fig:spec} show the \erosita\ spectra from two sky regions (eROSITA sky tiles 294147 and 085126, \citealt{Merloni24}) characteristics of bright and faint diffuse emission, respectively. 
The spectra shown in Fig.~\ref{fig:spec} here are meant to help the reader grasp the meaning and relations between different emission lines and the spectral components producing them.
Both spectra show several peaks associated with the most prominent transitions, which stand out in the \erosita\ spectra above the pseudo continuum (Fig. \ref{fig:spec}). 

We focus this work on the two strongest lines observed in the \erosita\ spectra of the diffuse emission: \os\ and \oe\ (see Tab. \ref{tab:line}).
We chose the width of the energy bands on the basis of the energy resolution of the pnCCD cameras of \erosita, as determined on ground \citep[see Table 4 in][]{Predehl2021}. 
Within these relatively narrow energy ranges, fainter emission lines might contribute to the emission from the pseudo continuum. For this reason, we did not perform any continuum subtraction of the emission lines (e.g., computing the line equivalent width). 
Instead, we forward modeled the expected emission (assuming the AtombDB library of atomic data and transitions \footnote{\url{http://atomdb.org/}}) in order to compare the data with expectations (see Sect.~\ref{sec:subtract}). 

\begin{figure*}
    \centering
    \includegraphics[width=0.9\textwidth,angle=0]{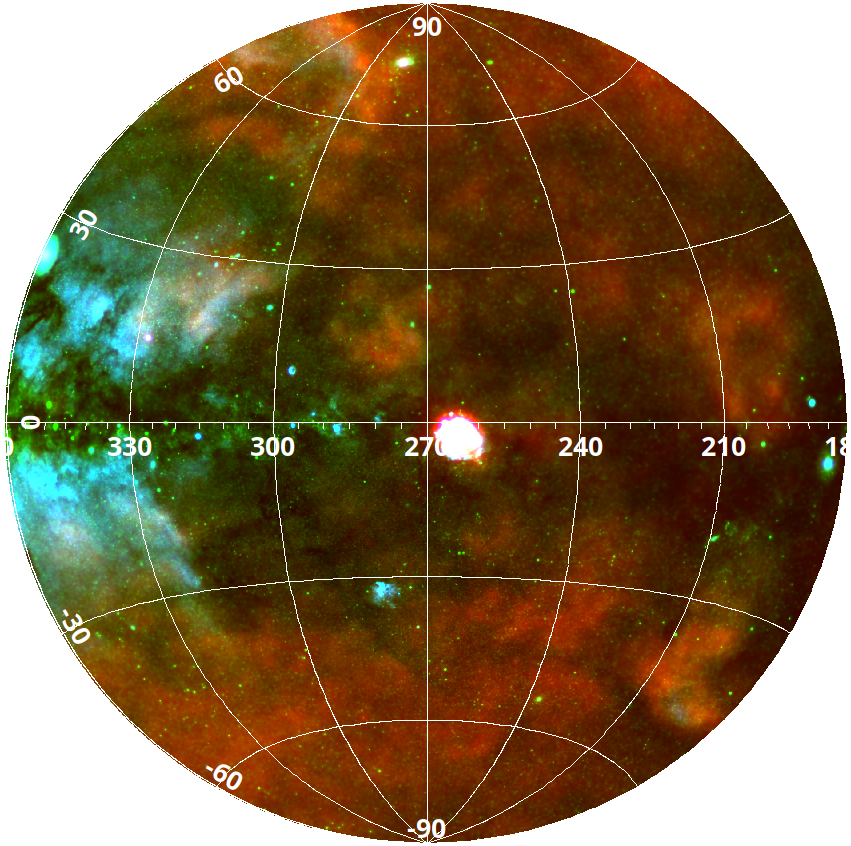}
    \caption{Bright emission lines in eRASS1. Thre RGB map is composed by the broadband maps at 0.20--0.25 keV (red, \citealt{Zheng2023}), 0.2--2.3 keV (green, \citealt{Zheng2023}) and the narrowband \oe{} (blue, this work) images. All the maps include the contribution of different components (both celestial and instrumental) and were smoothed with an adaptive kernel set to retrieve S/N threshold of 15.}
    \label{fig:RGB}
\end{figure*}

Figure \ref{fig:RGB} shows an RGB half-sky map of the emission in the \os, the broad continuum (0.2-2.3 keV) and \oe\ bands in red, green and blue, respectively. 
The RGB image shown in Fig.~\ref{fig:RGB} highlights the emission associated with the hot plasma. 
All major extended features which are also observed in the broadband maps are seen in Fig. \ref{fig:RGB} \citep[see][for a chart map of the principal diffuse structures]{Zheng2023}. 
In particular, the nearby extended features, such as the Monogem ring \citep{Knies2022}, the Antlia supernova remnant, and the Orion-Eridanus complex \citep{2019A&A...631A..52J} appear as large-scale extended features with a markedly orange colour. 
A greenish and bright patch of X-ray emission is observed around the LMC \citep{2024arXiv240117291L}. 
A greener-bluer color characterises the interior of the \erosita\ bubbles \citep{Predehl2020Natur} compared with the anti-center direction. 
Moreover, clear color gradients are observed at the edges of the \erosita\ bubbles. 
The dark regions at small Galactic latitudes $b$ (i.e. along the Galactic plane) in Fig. \ref{fig:RGB}, some of which are as extended as $\sim20$ deg, are the result of absorption of the soft X-ray emission through the ISM.

\begin{table}
\centering
\begin{tabular}{lccll}
Brightest       & Transition      & Theoretical          & Image energy     \\
line            & levels      & energy [eV]         & range [eV]      \\
\hline
O VII & $7\longrightarrow 1$  & 574              & 534--614   \\
O VIII & $4,3\longrightarrow 1$ & 654, 653           & 614--694   \\
\end{tabular}
\caption{Transitions considered in this work and corresponding energy band selected to extract photon lists and to create the corresponding images. }
\label{tab:line}
\end{table}

\subsection{The dependence of the ion fractions on temperature}

Figure \ref{fig:ionization_T} shows the expected ion fraction as a function of the plasma temperature for oxygen, as derived from AtombDB under the assumption of an optically thin hot plasma mainly ionized by collisions between atoms. 
For each element, the ionization fraction is obtained by balancing the net rates of ionization and recombination, which depends on the temperature of the plasma.
The vertical dotted grey lines indicate the typical or empirical temperature of some of the components contributing to the soft X-ray diffuse emission. 
These are: the local hot bubble (LHB), characterised by $T\sim0.097$ keV \citep{Liu2017ApJ}; the circumgalactic medium (CGM or halo) of the Milky Way, with $T\sim$0.15-0.2 keV \citep{Ponti2023}; the eROSITA bubbles, which we assume to have $T\sim0.3$ keV (\citealt{2018Galax...6...27K}; but see also \citealt{2023NatAs...7..799G}). 
Figure \ref{fig:ionization_T} shows that plasma with temperatures characteristic of the LHB is dominated by \os, while the CGM emission has similar contributions from \os\ and \oe, while the \oe/\os\ line ratio becomes insensitive to the presence of plasma at even higher temperatures (i.e., $T\gg0.3$~keV). 

\subsection{Soft X-ray line ratios as temperature diagnostics}
\label{sec:ratio}

\begin{figure}
    \centering
    \hspace{-0.7cm}
    \includegraphics[width=0.5\textwidth, clip]{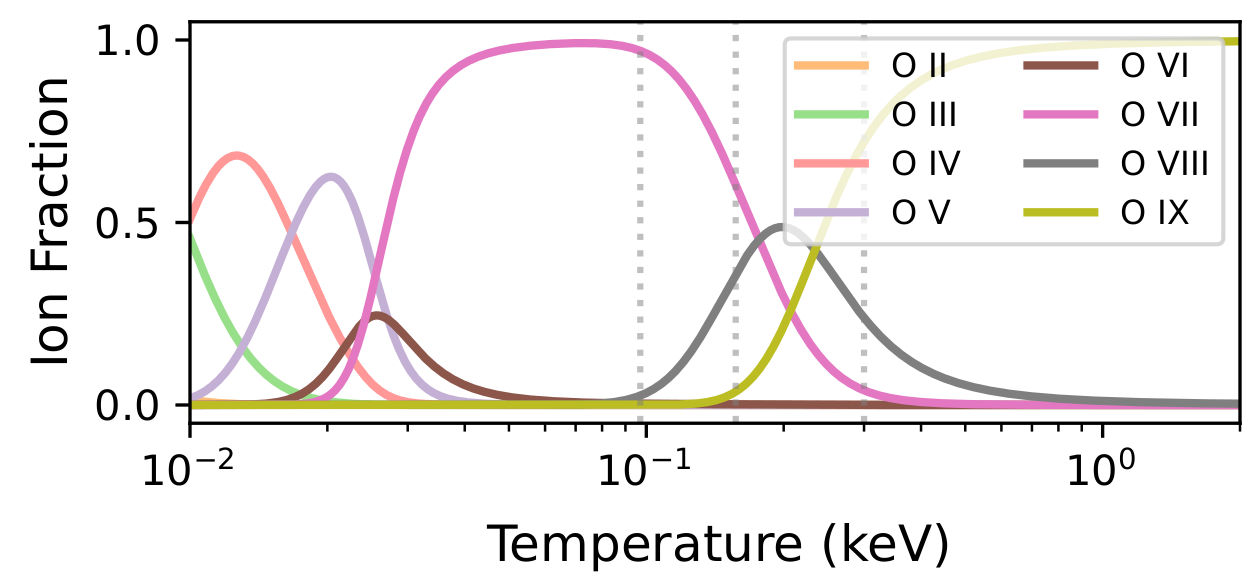}
    \caption{Ionization fraction of oxygen as a function of temperature \citep[from AtomDB, see ][]{2001ApJ...556L..91S}. The vertical dotted lines in gray show the typical temperature of: LHB ($T=0.097$ keV); CGM ($T=0.15$ keV); eROSITA bubbles (assumed to be $T=0.3$ keV).}
    \label{fig:ionization_T}
\end{figure}

As can be seen in Fig. \ref{fig:ionization_T}, under the assumption of a single emission component in collisional ionization equilibrium, it is possible to measure the temperature of the emitting plasma, given an observed line ratio. 
In particular, if both ions belong to the same element (e.g., \os\ and \oe), then the line ratio has almost no dependence on the metal abundances of the emitting plasma, therefore becoming a reliable tool to determine the temperature of the hot plasma. 

\begin{figure}
    \centering

    \includegraphics[width=0.5\textwidth]{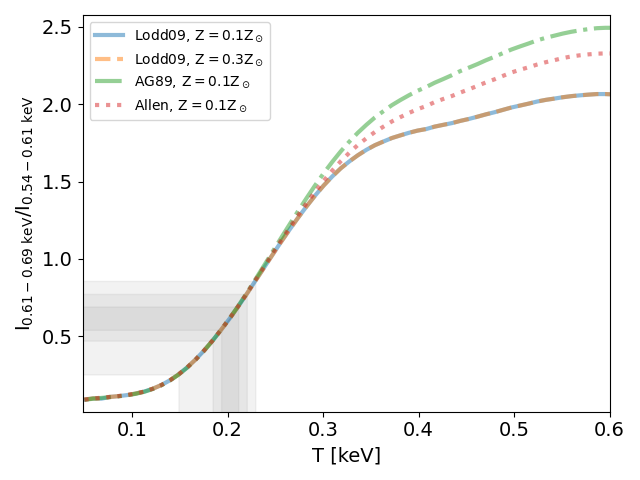}
    \caption{Narrowband intensity ratio as a function of temperature, absolute and relative metal abundances \citep[as labelled, from AtomDB][and convolved with the \erosita\ spectral resolution]{2001ApJ...556L..91S}. The shaded areas encompass the 2, 10, 25, 75, 90, 98\% of the distribution of the the narrowband intensity ratio analyzed in this work (an image of the data is later presented in the top panel of Fig.~\ref{fig:Tmaps_vs_absorption}).
    }
    \label{fig:plot_O8vsO7_vs_T_abundsets}
\end{figure}

Figure \ref{fig:plot_O8vsO7_vs_T_abundsets} shows the evolution of the \oe/\os\ line ratio as a function of temperature. 
The line ratio was computed by simulating a collisionally-ionized diffuse gas ({\sc apec} component in {\sc Xspec}) at each temperature. 
The theoretical spectrum obtained was then convolved with a Gaussian kernel with a full width at half maximum of 74 eV in order to simulate the spectral resolution of \erosita\ and compare the theoretical ratio with the data.
Finally, the fluxes in each band (see Tab.~\ref{tab:line}) were used to produce the expected line ratio as a function of temperature.
The lines in the bottom panel of Fig. \ref{fig:plot_O8vsO7_vs_T_abundsets} show that the \oe/\os\ line ratio as observed by \erosita\ is an excellent temperature diagnostic for single temperature plasma between $T\sim0.1$ and $T\sim0.3$~keV, while it is insensitive to plasma hotter than $T\geq0.3$~keV.
The different lines show line ratios under the assumption that the emitting plasma has a metallicity of 0.1 or 0.3 Solar, as well as holding different ratios between the abundances of the metals (e.g. assuming different abundance sets; \cite{2009LanB...4B..712L}, \cite{Anders1989GeCoA} and \cite{2019LPICo2189.2009A}).
As expected, in the 0.1-0.3 keV temperature range, where the \oe\ and \os\ lines are sufficiently bright with respect to the pseudo-continuum, little to no variations of the line ratio are observed with metallicity. 


\section{Subtraction of non-CGM emission components from the maps} 
\label{sec:subtract}

As discussed in Section \ref{sec:ratio}, from the line ratio maps it is possible to extract important information on the temperature of the CGM, if the emission can be described by a single component of an optically-thin plasma in collisional ionisation equilibrium. 
However, the diffuse soft X-ray emission measured by eROSITA has several contributions from various thermal and non-thermal components that need to be removed before we can constrain the temperature of the CGM. 
Additionally, we expect that the emission from the CGM is, at least in part, absorbed by the interstellar medium, further complicating the interpretation of the line ratio maps. We note that the subtraction of (back-)foreground components is not performed on all the images shown within this paper but is applied to produce and analyze the temperature distribution of some representative sky patches. Below, we provide a description of the corrections we applied to retrieve the temperature map from the data.

\subsection{Subtraction of the instrumental background}

During the eRASS1 survey, the level of the instrumental background has been frequently monitored, thanks to weekly observations with the filter wheel closed \citep{Yeung2023}. 
The analysis of these filter wheel closed data shows that the \erosita\ instrumental background has been relatively stable in flux and spectrum for the entire duration of eRASS1.

Such instrumental background is described by a relatively flat spectrum with various emission lines, as well as detector electronic noise, which is stronger at lower energies and can be represented by a rise in the flux of the instrumental background towards lower energies \citep[solid black line in Fig.\ref{fig:spec}; see also][]{Yeung2023}. 
Here, we subtracted the contribution of the instrument background over the entire sky by assuming that it is stable over the duration of the eRASS1 survey. 

We stress that the contribution of the background over the narrow \os\ band is small, being $\sim$40 and $\sim$10 times fainter than the O~{\sc vii} emission in the bright and faint region we denoted in Fig.\ref{fig:spec}, respectively. 

\begin{figure}
    \centering
    \includegraphics[width=0.49\textwidth, clip, trim=1.0cm 1.0cm 1.8cm 1cm]{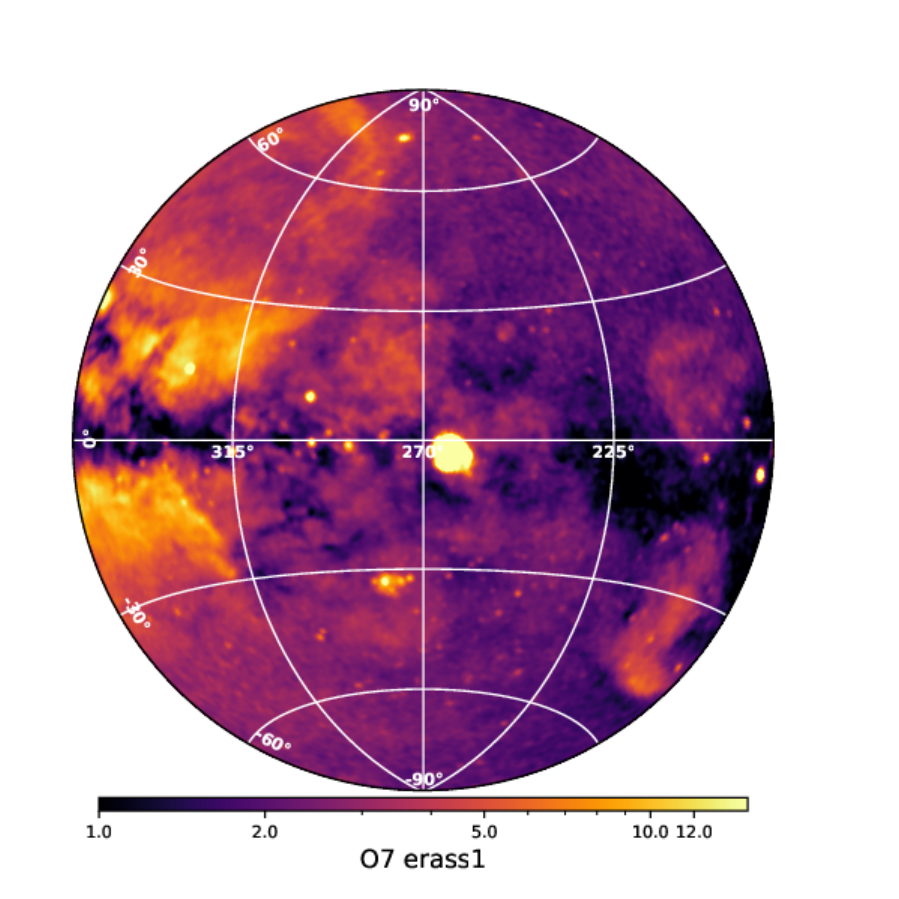}
    \caption{Narrowband map in the energy range of the \os{} line as observed by \erosita\ during eRASS1 in units of $\rm cts\, s^{-1}\, deg^{-2}$. An adaptive smoothing kernel using S/N=15 was applied.}
    \label{fig:os}
\end{figure}
\begin{figure}
    \centering
    \includegraphics[width=0.49\textwidth, clip, trim=1.2cm 1.0cm 1.7cm 1cm]{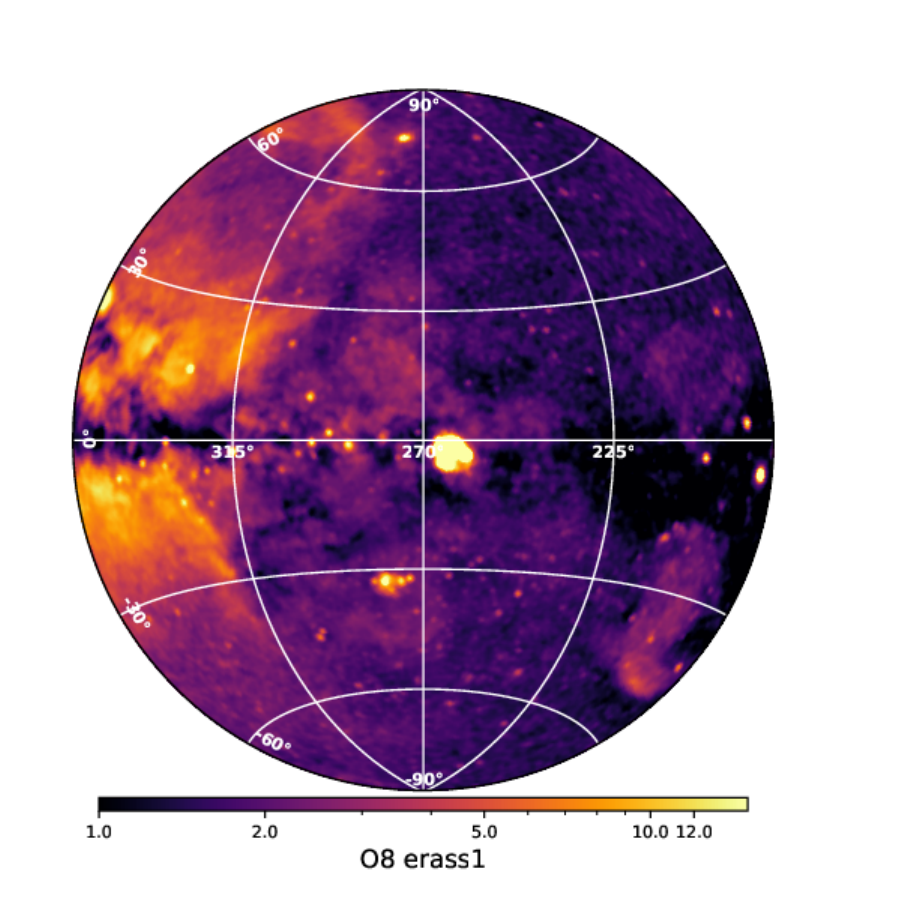}
    \caption{Narrowband map in the energy range of the \oe{} line as observed by \erosita\ during eRASS1 in units of $\rm cts\, s^{-1}\, deg^{-2}$. An adaptive smoothing kernel using S/N=15 was applied. }
    \label{fig:oe}
\end{figure}
\subsection{Removing the contribution from the solar wind charge exchange emission}

eRASS1 occurred close to solar minimum when the heliospheric solar wind charge exchange (SWCX) is at its minimum \citep{Kuntz19}. In addition, \erosita\ is located close to the Sun-Earth Lagrangian point L2 (where the effects of the geocoronal SWCX are low). 
In this context, \citet{Ponti2023} and \citet{Yeung2023} have reported evidence for a time-variable component affecting the diffuse emission observed by \erosita. 
This variable component was attributed to the effects of the heliospheric SWCX. 
The impact of SWCX is found to be minimal during eRASS1 \citep{Yeung2023, Ponti2023} due to a combination of the aforementioned effects: the telescope position at L2 and eRASS1 happening close to solar minimum.
The contribution from time-variable components in the sky (including and setting an upper limit to the SWCX) was computed by first producing background maps reporting the minimum values of \os{} and \oe{} observed by \erosita{} between 2019 and 2022 in every direction in the sky. This is possible thanks to the multiple passages of the telescope's field of view, according to the survey scanning strategy. The minimum map was then used to detect any variable excess in time. We then subtracted the map representing the time-variable excess from the narrowband maps of Figs.~\ref{fig:os} and ~\ref{fig:oe}.

\subsection{Removing the contribution from the LHB}

The LHB is an irregular volume filled with hot ($T \sim 0.097$ keV) plasma with a typical size of $\sim$200 pc around the Sun \citep[see e.g.][and references therein]{Yeung2023}. 
Being almost completely devoid of cold material, the emission from the LHB appears unaffected by absorption of X-rays. 

Because of its unabsorbed nature and relatively low temperature, the LHB emission dominates mainly in the very soft X-ray band ($\leq0.3$ keV).
\cite{Zheng2023} have shown that the diffuse emission observed by \erosita\ in the softer bands (0.2-0.25 keV) is indeed heavily affected by the emission from the LHB. 
For any study of the Galactic plasma beyond of LHB, such as the CGM and the \erosita\ bubbles, it is therefore necessary to remove the foreground contribution from the LHB.

The distribution of the LHB is observed to be inhomogeneous, based on the thermal emission from the LHB detected by ROSAT{} in the softest energy bands \citep{Snowden1997ApJ}.

Thanks to relatively recent observations from the Diffuse X-rays from the Local Galaxy mission (DXL), \cite{Liu2017ApJ} modelled the SWCX contribution in the ROSAT R12 map and inferred the latest 3D structures of the LHB.
We used the emission measure of the LHB provided in their work, assuming a constant temperature ($T=0.097$ keV {\sc apec} model with \citet{2003ApJ...591.1220L} abundance table) to reconstruct the observed LHB emission as observed by \erosita\ in each relevant energy band. We then removed the LHB component from the maps. 

Because of the relatively low temperature of LHB, the soft band is affected more than the harder bands. 
Figure 5 of \citealt{Ponti2023} shows that the LHB generally takes a much larger fraction in the \os\ narrow bands flux than that in \oe: in most regions of \os\ maps, the LHB takes more than 15\% of the flux, while in \oe\ maps the contribution from LHB is smaller by about a factor 10.

\subsection{Absorption correction}

The soft X-ray spectrum is heavily affected by absorption from neutral material \citep{Locatelli2022}. 
To account for the absorption effect, we used the column density maps combining the HI \citep{HI4PI2016AA} and H$_2$ column density derived from Planck observations \citep{2016A&A...596A.109P, 2018JOSS....3..695M} following the recipe proposed by \cite{2013MNRAS.431..394W}.  
Then, assuming Solar abundances for the absorbing medium, we derived the correction factor to be applied to de-absorb either the maps or the emission from the various emission components \citep[see Appendix A in][for further details]{2023arXiv231010715L}. 
The absorption correction was applied to the CGM and CXB components only, because we assumed that the SWCX and LHB components are un-absorbed. We note that the correction for the effects of absorption are negligible for column densities lower than $N_H\leq10^{21}$~cm$^{-2}$ in the oxygen maps. This condition applies to all the regions included in our analysis of the temperature distribution.
\begin{figure*}
    \centering
    \includegraphics[width=0.9\textwidth]{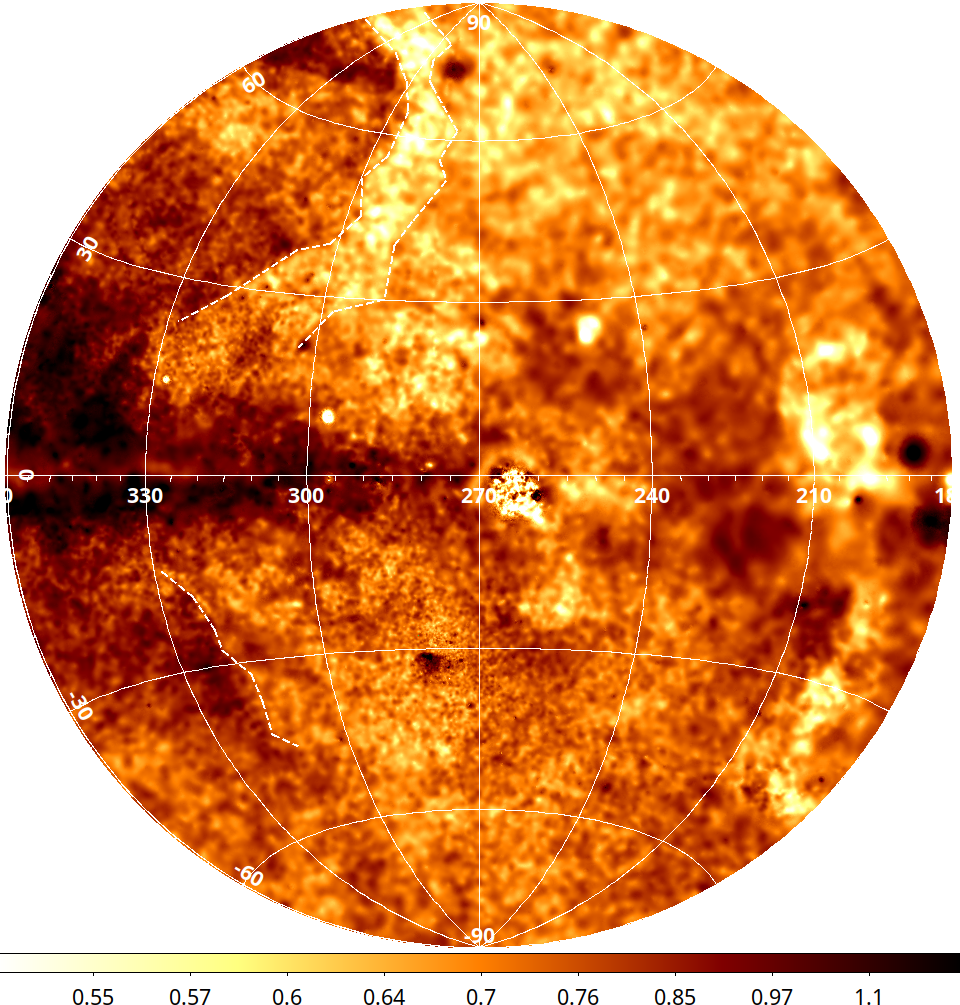}
    \caption{\oe/\os\ line intensity ratio map (unitless), as observed by \erosita\ during eRASS1. The contribution from CXB, LHB, foreground absorption and instrumental background (see Sect.~\ref{sec:subtract}) has not been removed before taking the line ratio shown in this image. An adaptive smoothing kernel using S/N=15 was applied. The kernel used was the same for the \os{} and \oe{} maps. The white dashed lines are placed along the sharpest features close to the boundaries of the \erosita\ bubbles. }
    \label{fig:RatO}
\end{figure*}
\subsection{Subtraction of the cosmic X-ray background (CXB)}\label{sec:cxb}

The major contribution to the diffuse X-ray emission above $\sim0.6$~keV is given by the CXB \citep[see Tab. 4 of][]{Ponti2023}. 
Ultra-deep X-ray surveys performed with Chandra and XMM-Newton\ have resolved more than $\sim80$ \% and $\sim92$ \% of the CXB flux into extra-galactic discrete sources (primarily active galactic nuclei, clusters of galaxies, groups, normal galaxies, etc.) in the 0.5--2 and 2--7 keV bands, respectively \citep{2017ApJS..228....2L, 2021AAS...23820901B}. 
The CXB flux is observed to be uniformly distributed, with small amplitude variations on arcminutes to degrees scales. 
We removed the CXB component by assuming it has constant flux on the sky, that it is absorbed by the full Galactic column density of material (computed as above), and that its spectrum can be described by a doubly broken power law \citep{Ponti2023}. 
The CXB spectrum \citep{Gilli2007AA} is assumed to have a photon index of $\Gamma_{3}=1.45$ above 1.2 keV, of $\Gamma_{2}=1.6$ between 0.4 and 1.2 keV and $\Gamma_{1}=1.9$ below 0.4 keV and to have a normalisation of 8.2 photons s$^{-1}$ cm$^{-2}$ at 1 keV. 

\section{The \os\ and \oe\ line maps of the CGM component}
\label{sec:narrowband_maps}

Figure \ref{fig:os} shows the half sky emission observed by \erosita\ during eRASS1\ within the \os\ energy band, while Figure \ref{fig:oe} shows the same map in the \oe\ band. 
In both bands bright emission is observed at relatively high Galactic latitudes $|b|\geq30^\circ$, with count rates of $\sim0.01$ cts~s$^{-1}$ deg$^{-2}$. 
Some small depressions are observed there, however they are associated with absorbing clouds \citep{2015ApJ...806..120S, Yeung2023, 2023A&A...670A..99P}. 
This indicates that, as expected, we are surrounded by hot plasma along every direction on the sky, the so called hot CGM. 

As already pointed out, along most of the Galactic disc, the effect of interstellar absorption is much more relevant. 
Indeed, dark patches are observed nearby dark clouds, which are primarily concentrated at small latitudes. 
\os{} and \oe{} are the brightest lines observed by \erosita\ in the soft band. For this reason many of the features described in both the broadband maps \citep{Zheng2023} and in the RGB map in Fig.~\ref{fig:RGB} are also evident in the narrowband maps of the oxygen lines. For instance, large-scale enhanced \os\ and \oe\ emission is observed both in the direction towards the Galactic center, along the footprint of the \erosita\ bubbles and on smaller regions towards the anti-center, (e.g. Orion-Eridanus superbubble; Monogem ring; Antlia SNR).

It is also possible to observe a clear trend of increasing \oe\ emission closer to the Galactic disc and towards the Galactic center (Fig. \ref{fig:oe}). These large-scale features contain morphological information recently used to constrain the geometry and physical properties of the warm-hot CGM \citep{2023arXiv231010715L}.
\section{Line ratio and pseudo-temperature maps} \label{sec:ratios}

Figure \ref{fig:RatO} shows the ratio of the \oe/\os\ emission. 
The \os\, and \oe\, maps used to compute the ratio collect counts over a squared pixel of side length $\rm \Delta r_{pxl}=2 deg$. This pixel size allows most pixels in the maps to hold $\rm C \geq 30$ counts. Under the $C\gg 1$ assumption, the uncertainty related to the count can then be computed as $\sqrt{C}$. We note that the only regions where the assumption is broken are those highly absorbed, primarily along the galactic disk and towards dark clouds \citep{Yeung2023}. The galactic disk regions are anyway not included in the quantitative analysis, and the results are presented in the following sections. 
We stress that, for display purposes, we have inverted the color scale, so that brighter regions correspond to a higher contribution from \os, therefore to a lower temperature. 
At high Galactic latitudes $|b|>30^\circ$ and away from the Galactic center $l<270^\circ$, the line ratio map appears rather smooth (with small fluctuations, that do not appear significant, see Sect.~\ref{sec:CGM_T}), with the northern hemisphere possessing slightly lower ratio. 
On the contrary, at high Galactic latitudes $|b|>30^\circ$ towards the Galactic center $l<270^\circ$, sharp color variations are clearly observed in Fig. \ref{fig:RatO}. 
Particularly evident is a bright (less hot) stripe running all along the edge of the northern \erosita\ bubble (enclosed by the dashed white lines drawn in Fig.~\ref{fig:RatO}). 
Just inside (closer to the Galactic center) this bright stripe, a dark (hot) region is observed to run again all along the edge of the northern \erosita\ bubble. 
Despite fainter, such dark (hot) transition layer can be observed also along a good fraction of the southern \erosita\ bubble. 

\begin{figure*}
    \centering
    \includegraphics[width=0.9\textwidth,angle=0]{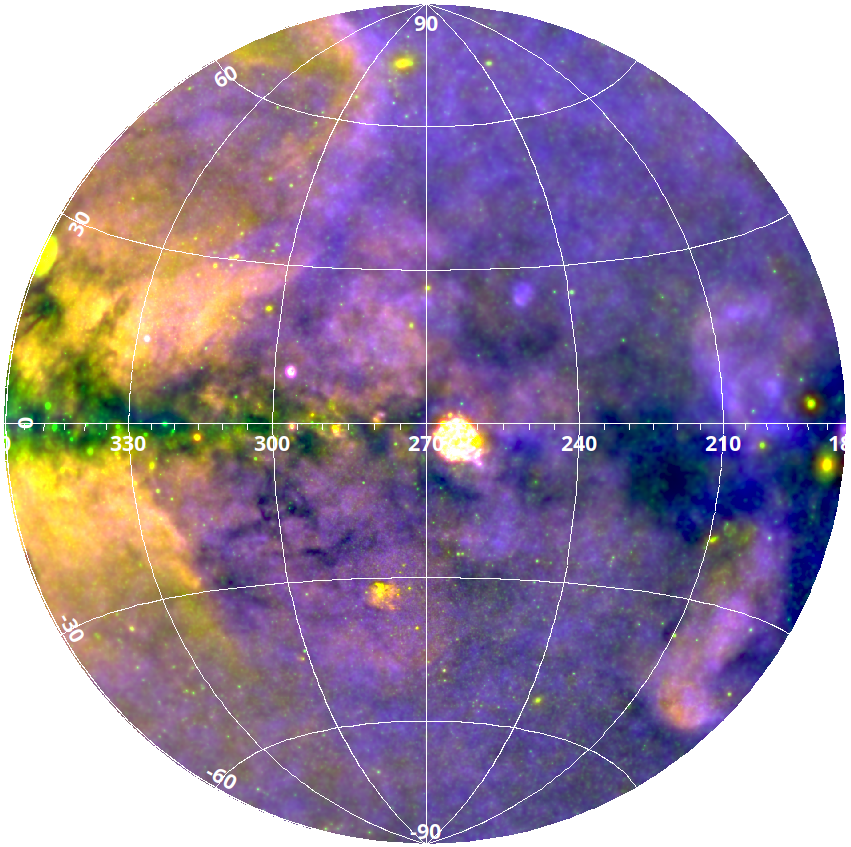}
    \caption{Bright emission lines in eRASS1. The RGB map is composed of \os{} (red), broadband emission in the 0.2--2.3 keV energy range (green) and \os{} to \oe{} line ratio (blue). All the maps include the contribution of different components (both celestial and instrumental) and were smoothed with an adaptive kernel set to retrieve S/N threshold of 15.}
    \label{fig:RGB2}
\end{figure*}
To better highlight this interesting feature occurring at the edge of the \erosita\ bubbles, Fig.~\ref{fig:RGB2} shows both the X-ray continuum (0.2-2.3~keV) in green as well as the \os{} to \oe{} line ratio in blue and the \os\ emission in red. 
The \os{} to \oe{} line ratio in blue is thus brighter whenever a source produces more \os{} than \oe{}, that is for lower plasma temperatures (see also Fig.~\ref{fig:ionization_T}). Fig.~\ref{fig:RGB2} thus clearly shows that this colder rim of emission occurs in close correspondence with the surface brightness transition used to define the edges of the \erosita\ bubbles.
This effect is very clear along the northern bubble but it can be followed all the way down to about $-40^\circ$ also in the southern bubble (Fig. \ref{fig:RGB2} and \ref{fig:Tmaps_vs_absorption}), although partially blended with broad absorption features close to the boundary of the bubble, in projection.

\subsection*{Pseudo-temperature map derived from the \oe/\os\ line ratio}
\label{sec:kTO}

\begin{figure}
    \centering
    \includegraphics[width=0.472\textwidth]{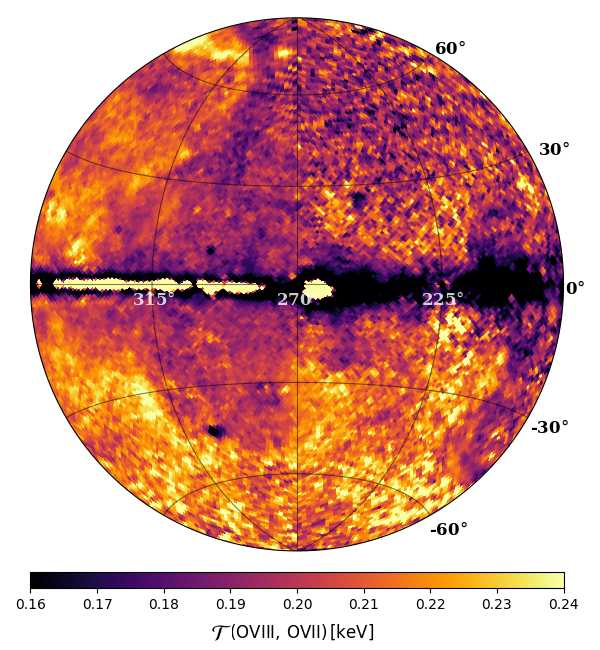}
    \includegraphics[width=0.472\textwidth]{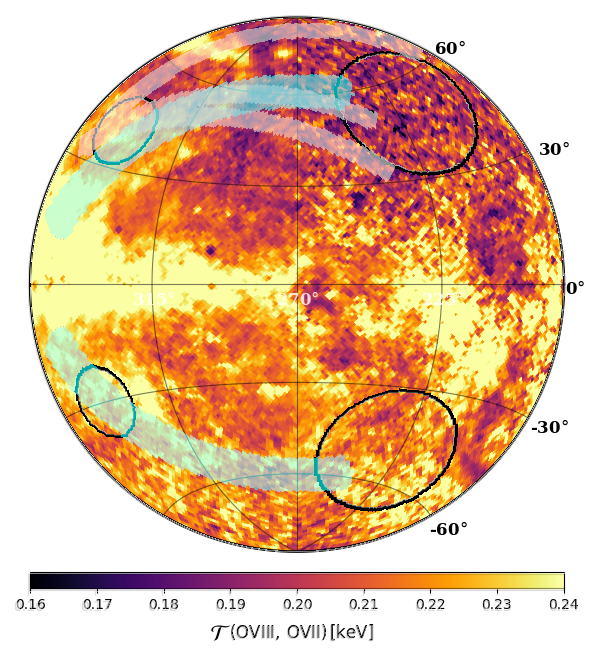}
    \caption{Pseudo-Temperature derived from O lines ratio with(out) accounting for the absorbed intensity of the line. Upper panel: temperature map corrected for foreground absorption. This version is used for quantitative statements on the temperature throughout the paper. These maps were corrected for the LHB, CXB, SWCX, instrumental background emissions and foreground absorption, as presented in Sect.~\ref{sec:subtract} (while in Fig.~\ref{fig:RatO}, the ratio was not corrected). The regions used for the pseudo-temperature analysis are shown in the panel below. Lower panel: temperature map, without deabsorbing the \os{} and \oe{} line emission. The black circles show the regions selected for the analysis of the \erosita\ bubbles and the CGM in the north and south hemispheres. The profiles shown in Fig.~\ref{fig:plot_profile_eRObubble_T_NvsS} and Fig.~\ref{fig:plot_profile_eRObubbleN_paths123} are indicated by the transparent cyan and white regions, respectively. }
    \label{fig:Tmaps_vs_absorption}
\end{figure}

Assuming that the line ratio map shown in Fig. \ref{fig:RatO} is produced by only one warm-hot plasma component, by inverting the relation in Fig.~\ref{fig:plot_O8vsO7_vs_T_abundsets}, from the oxygen lines ratio map, we are able to create the first temperature map of the warm-hot CGM of the Milky Way (see Fig.~\ref{fig:Tmaps_vs_absorption}, top panel). 
Of course, such a "temperature map" is valid only in areas where a single plasma component is emitting most of the observed photons. 
In other regions, where different components of comparable brightness build up the observed narrowband intensities, the derived map loses its meaning as a temperature. 
For this reason, a meaningful temperature map depends on an accurate selection of the emission component subject of the study, for instance, through the models and subtractions presented in Sect.~\ref{sec:subtract}. 
Additionally, absorption can severely affect the temperature map by absorbing the emission of the different lines by a different factor, therefore affecting the final temperature map. 
Because in several places, the assumption of a single thermal component is not true (e.g., within the footprints of the \erosita\ bubbles and close to the Galactic disc, where the hot ISM could contribute to the emission), we call the map derived from the oxygen line ratio as "pseudo-temperature" map. 

The top panel of Fig. \ref{fig:Tmaps_vs_absorption} shows the pseudo-temperature $\mathcal{T}$ map obtained by subtracting all components as detailed in Sect.~\ref{sec:subtract}. 
Therefore, all components discussed by \cite{Ponti2023}, apart from the CGM (composed by a hot halo, a disc-like component, and the emission of the Galactic outflow) have been removed. 
By looking at Fig.~\ref{fig:spec} (see also Fig. 5 of \cite{Ponti2023}), it is possible to see that, at moderate latitudes, little contribution is given to the \os\ and \oe\ lines by the emission attributed to the corona. 
This suggests that the pseudo-temperature map derived from the \oe\ over \os\ maps is a good proxy of the real CGM plasma temperature in regions away from the Galactic outflow and the Galactic disc. 
On the other hand, we expect that the additional emission induced by the Galactic outflow and the hot interstellar medium will break the initial assumption that only one thermal component is contributing to the oxygen line ratio map, therefore biasing the temperature map shown in Fig. \ref{fig:Tmaps_vs_absorption} within the footprints of the \erosita\ bubbles and along the Galactic disc. 

The top panel of Fig. \ref{fig:Tmaps_vs_absorption} shows similar structures to the ones observed in the oxygen lines ratio. 
In particular, the CGM pseudo-temperature appears rather uniform towards both the northern and southern hemispheres, with values in the order of $\mathcal{T}\sim0.20-0.22$~keV, respectively. 
A low-temperature shell (colder than the CGM) surrounds all of the northern bubble and a good fraction of the southern bubble. 
Such "shell" has a considerable width, appearing as extended as $\sim10^\circ$ in several places. 
We note that, just inside this colder shell, the temperature rises again to temperatures higher than the surrounding CGM. 
Again, this interior and hotter shell surround all of the northern bubble as well as a good fraction of the southern one (Fig. \ref{fig:Tmaps_vs_absorption}). 

Can the LHB alone be attributed to the \oe{}/\os{} line-ratio transition at the eROSITA bubbles? A 0.1\,keV LHB can be approximated to produce only \os{} but not \oe{}. Assuming a single-temperature CGM that gives a uniform \oe{}/\os{} line ratio across the sky, adding such a LHB component always decreases the line ratio, more at locations where the LHB emission measure is high. Since the \citet{Liu2017ApJ} LHB model shows a lower emission measure at $l\gtrsim300\degr$ (near eROSITA bubbles) compared to the rest of the western Galactic hemisphere, one should see a lower \os{} (hence higher \oe{}/\os{}) in the eROSITA bubbles compared to outside. While we model the LHB emission according to the best available information, deviations of $\pm 15\%$ from a uniform LHB temperature have been recently found \citep{Yeung2023} in a few directions towards giant molecular clouds, at odds with the model used in this work. Further improvements in the knowledge of the LHB thermodynamics will be beneficial to the analysis presented in this work.

In addition, we emphasize that our results and interpretations concerning the pseudo-temperature rely on the assumption of a single-temperature, collisionally-ionized plasma. In non-equilibrium scenarios, the observed lowering of the line ratio in the shell could be due to an under-ionization state rather than a lower temperature. As the ionization properties depend on the emission model, we present our results only based on the stated assumptions.

The bottom panel of Fig. \ref{fig:Tmaps_vs_absorption} shows how the temperature map changes once the correction for absorption is not applied. 
As expected, only in highly absorbed regions (i.e., along the Galactic disc and towards dark clouds) do the corrections become relevant.

\section{Constraining the temperature distribution of the CGM} \label{sec:CGM_T}

\subsection{On the CGM temperature distribution on small scales (few-to-ten degrees)} 
\begin{figure}
    \centering
    \includegraphics[width=0.49\textwidth]{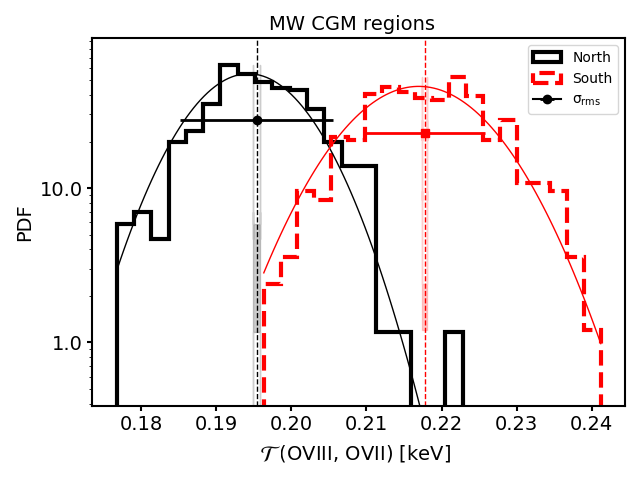}
    \caption{Probability distribution function (PDF) of the pseudo-temperature derived in the north and south CGM regions (black solid and red dashed lines, respectively; see footprints in the lower panel of Fig.~\ref{fig:Tmaps_vs_absorption}). The thin solid lines show Gaussian fits of the temperature distributions. The vertical dashed lines show the median values of the distribution with the uncertainty on the mean bracketed by the filled vertical regions.
    }
    \label{fig:hist_T_NvsS}
\end{figure}
To constrain the pseudo-temperature distribution of the CGM away from the Galactic disc and center, we selected two circular regions centered at (l,b) = (220,+50) and (l,b) = (230, -50) and with radii of 20 degrees (black solid wide circles in Fig.~\ref{fig:Tmaps_vs_absorption}, bottom panel). 
Two additional regions at coordinates (l,b) = (340,+40) and (l,b) = (340, -30) with radii of 10 degrees were also considered to study the pseudo-temperature within the \erosita\ bubbles (black solid small circles in Fig.~\ref{fig:Tmaps_vs_absorption}, bottom panel).

Choosing moderate- and high-latitude regions minimizes potential systematic uncertainties in the oxygen ratio introduced by the absorption model used to retrieve the unabsorbed CGM emission in both oxygen bands (see Appendix~\ref{sec:absorption}). 
Figure \ref{fig:hist_T_NvsS} shows the probability distribution function of the pseudo-temperature within the selected regions, obtained using a pixel side length of $\rm \Delta r_{pxl} = 2$ deg.  The distributions from the northern and southern hemispheres are shown by the black solid and red dashed lines, respectively. In Fig.~\ref{fig:hist_T_NvsS}, the horizontal error bars show the root-mean-square (rms) value $\rm \sigma_{rms}$ of the pseudo-temperature uncertainty in the regions considered. 

The value $\rm \sigma_{rms}$ happens to be similar to the standard deviation $\rm std(\mathcal{T})$ of the pseudo-temperature map computed in the same region (i.e. $\rm \sigma_{rms} \simeq std(\mathcal{T})$), in both the north and south regions. 
We conclude that the CGM pseudo-temperature can be considered uniform over the regions considered, that is, between 2 and 20 deg scales, given the statistical uncertainties on the pseudo-temperature values. 
Any intrinsic fluctuation $\Delta T$ would, in fact contribute to the overall spread of the distribution $\rm std(\mathcal{T})$ following \citep[e.g.][]{2003MNRAS.345.1271V}:
\begin{equation}
    \rm \left( std(\mathcal{T}) \right)^2 = (\Delta T)^2 + \sigma_{rms}^2. \label{eq:signal}
\end{equation}
Therefore, provided $\rm std(\mathcal{T})\simeq \sigma_{rms}$, we constrain any intrinsic pseudo-temperature fluctuation to be $\rm \Delta T\leq \Delta \mathcal{T} \leq \sigma_{rms} = 0.007$ keV on $\Delta \theta \simeq 2-20$ deg in the CGM regions. 
The fractional fluctuation with respect to the average pseudo-temperature $\Delta \mathcal{T} / \langle \mathcal{T} \rangle$ (reported below in Sect.\ref{sec:asymmetry}) amounts to $\Delta \mathcal{T} / \langle \mathcal{T} \rangle \leq 4\%$ for both the north and south regions.
In particular, in the south region we find $\Delta \mathcal{T} = 0.0062$ keV (i.e. $\Delta \mathcal{T}>0$), hinting to a detected fluctuation of the pseudo-temperature of $\Delta \mathcal{T} / \langle \mathcal{T} \rangle = 2.7 \%$. Due to the \erosita{} scanning strategy, the southern hemisphere is characterized by a higher exposure than the north, thus providing relatively deeper data. 

We assessed the statistical uncertainty on the potential fluctuation by exploiting both a jackknife resampling and a bootstrap of the pseudo-temperature distribution. The derived statistics on the simulated distributions are consistent between the two methods. We obtain that the values $\rm std(\mathcal{T}) \, and\, \sigma_{rms}$ of the simulated distributions vary (one standard deviation) by $3.7\%$ and $1.4\%$ of their expected value, respectively. We derive $\Delta \mathcal{T} = 0.0062 \pm 0.0004$ keV (statistical) by propagating the uncertainties.
Additional systematic biases however, may be introduced by the modeling of the back- and foreground components subtracted to retrieve the CGM intensity. By switching on and off the subtraction of the different components (i.e. CXB, LHB, SWCX) and increasing their flux by an arbitrary $10\%$, we find a shift in the detected excess of $\pm 0.0013$ keV. The uncertainty on the fluctuations is then dominated by the systematic component. We obtain $\Delta \mathcal{T} = 0.0062 \pm 0.0004$ keV (statistical) $\pm 0.0013$ keV (systematic), translating to $\Delta \mathcal{T} / \langle \mathcal{T} \rangle = 2.7 \% \pm 0.2\%$ (statistical) $\pm 0.6\%$ (systematic).
The detected fluctuation thus seems significant (with respect to zero). We only investigated systematic uncertainties by modifying the relative importance of back- and foreground components without changing their morphology. For this reason, we are still cautious about the source of this signal. Further investigations exploiting the full amount of the \erosita{} data (ongoing) will be able to confirm or disprove the origin of this excess in the scatter of the line ratio distribution and better assess which fraction of the observed fluctuation has to be ascribed to other components. This measure however, sets a tight upper limit to the temperature variations of the warm-hot CGM.

\subsection{North-south asymmetry in the pseudo-temperature distribution of the CGM} 
\label{sec:asymmetry}

As already pointed out when describing Fig. \ref{fig:RatO}, we note that the CGM in the southern hemisphere appears hotter (i.e. higher line ratio) than its northern counterpart.
From the oxygen pseudo-temperature map, the derived mean pseudo-temperature values are $\langle \mathcal{T} \rangle = 0.195\pm 0.001$ keV and $\langle \mathcal{T} \rangle = 0.218\pm 0.001$ keV for the north and south regions respectively. The quoted uncertainties indicate the error on the mean. The median pseudo-temperatures are equal to the mean pseudo-temperatures within uncertainties. The difference between north and south pseudo-temperature is thus $\Delta \mathcal{T} = 0.024\pm 0.001$ keV, where the uncertainty has been obtained from the quadrature of the mean pseudo-temperature uncertainties of the north and south samples. Thus, we find a significant difference of $\Delta \mathcal{T} / \langle \mathcal{T} \rangle \simeq 12\%$ in the pseudo-temperature of the hot CGM when comparing representative regions in the north and south Galactic hemispheres. This difference is larger than the sum of statistical and systematic uncertainties (see Appendix~\ref{sec:absorption}), and we thus interpret it as a physical, real effect. 

This then suggests that the pseudo-temperature of the CGM has variations as large as $\sim12$~\% on scales of several tens of degrees. In contrast, when looking at angular scales of the order of $\sim2-20^\circ$, then the pseudo-temperature distribution is consistent with being normal and shows variations as small as $2-3\%$, or in any case smaller than $\Delta \mathcal{T}\leq 4$~\%. 

\section{Constraining the pseudo-temperature distribution within the \erosita\ bubbles} \label{sec:bubbles_T}

\begin{figure}
    \centering
    \includegraphics[width=0.49\textwidth]{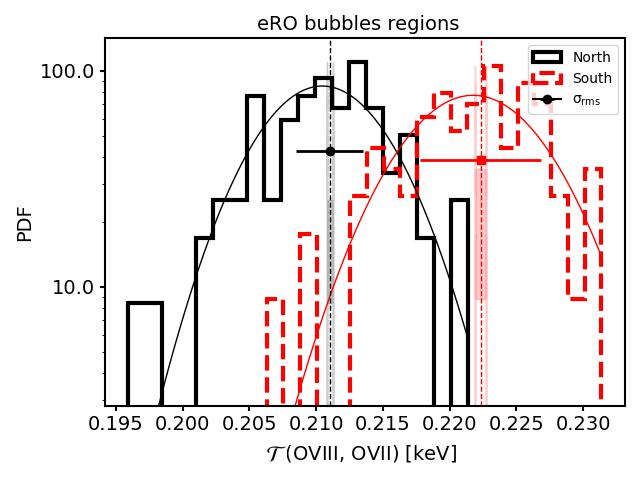}
    \caption{Probability distribution function (PDF) of the pseudo-temperature in the north and south regions inside the eROSITA bubbles (black solid and red dashed lines, respectively; see footprints in the lower panel of Fig.~\ref{fig:Tmaps_vs_absorption}). 
    }
    \label{fig:hist_eROB}
\end{figure}
Figure \ref{fig:hist_eROB} shows the probability distribution function of the pseudo-temperature within two circular regions inside the northern and southern \erosita\ bubbles with black and red colours, respectively. 
We stress again that, within this region, at least two thermal components (i.e., the emission from the \erosita\ bubbles, plus the emission from the other parts of the CGM) contribute to the total emission. Therefore, the pseudo-temperature derived from the line ratio must be biased. 
In particular, as shown in Figs. \ref{fig:ionization_T} and \ref{fig:plot_O8vsO7_vs_T_abundsets}, the pseudo-temperatures derived from the oxygen line ratio are unable to trace plasma hotter than $T\sim0.3$~keV. 
Indeed, the pseudo-temperatures derived from the oxygen line ratio are observed to be $\rm \langle \mathcal{T} \rangle = 0.211\pm 0.001$ keV and $\rm \langle \mathcal{T} \rangle = 0.217\pm 0.001$ keV for the northern and southern bubbles, respectively. 
These values are surprisingly similar to the temperatures derived for the CGM outside the \erosita\ bubbles, indicating that the oxygen line ratio is not sensitive to the presence of a hotter component within the bubbles (see Fig. \ref{fig:spec}). 
Therefore, despite the absolute value of the pseudo-temperature might be biased inside the \erosita\ bubbles (because the presence of a hotter component is not traced), we expect that intrinsic variations of the temperature of the \erosita{} bubbles would induce a spread in the distribution of oxygen line ratio and, consequently, on the pseudo-temperatures derived from the latter. 

Figure \ref{fig:hist_eROB} shows that the mean uncertainty on the value of the pseudo-temperature is significantly smaller than the pseudo-temperature standard deviation $\rm \sigma_{rms} < std(\mathcal{T})$, indicating that there are significant pseudo-temperature variations observed. 
The typical error bars within these regions inside the \erosita\ bubbles are smaller than the CGM regions, thanks to the brighter emission characterizing the bubbles compared to the background (faint) CGM. 
This suggests significant temperature variations are observed on small ($\sim2-10$ degrees) scales within the \erosita\ bubbles. 
We obtain $\rm std(\mathcal{T}) = 0.007$ keV in both regions and $\rm \sigma_{rms} = 0.003, 0.005$ keV in the north and south region respectively. From eq.~\ref{eq:signal}, we derive $\Delta \mathcal{T} = 0.006, 0.004$ keV. Compared to the average values $\langle \mathcal{T} \rangle = 0.2262 \pm 0.0007$ keV and $\langle \mathcal{T} \rangle = 0.2340 \pm 0.0007$ keV in the north and south regions within the \erosita{} bubbles, the relative strength of the excess scatter is $\Delta \mathcal{T} / \langle \mathcal{T} \rangle = 2.7, 1.9 \%$, respectively. The relative excess scatter thus shows similar strength inside and outside (see Sec.~\ref{sec:CGM_T}) the \erosita{} bubbles.

Potential deviations from a normal distribution (solid lines in Fig.~\ref{fig:hist_T_NvsS}) has been tested through a Kolmogorov-Smirnov (KS) test on both the north and south samples. 
The tested hypothesis is that of an underlying normal distribution of the sample. 
Both $\mathcal{T}$ and $\log \mathcal{T}$ distributions have been tested using the KS. 
The test hypothesis is considered as rejected for $p<0.05$. 
The resulting p-values do not reject the null hypothesis in all cases. 
The pseudo-temperature distributions are thus consistent with a normal distribution. 
A fit using a normal distribution is also shown (solid lines in Fig.~\ref{fig:hist_T_NvsS}).

\subsection{Pseudo-temperature variations across the \erosita\ bubbles} \label{sec:bubble_profiles}

In Sect.~\ref{sec:ratios}, we discussed the presence of pseudo-temperature variations across the edges of the \erosita\ bubbles. 
\begin{figure}
    \centering
    \includegraphics[width=0.49\textwidth]{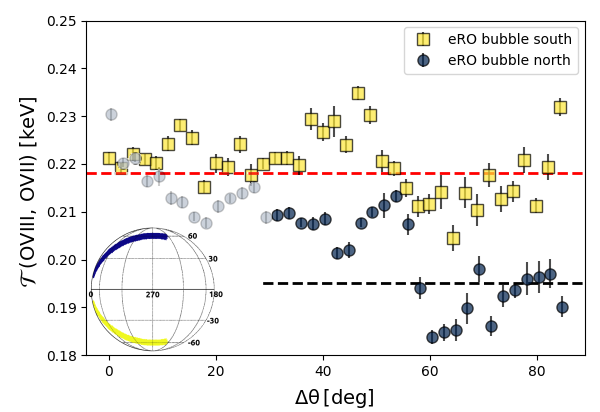}
    \caption{Pseudo-temperature profiles across the \erosita\ bubbles. The paths are also shown as cyan transparent regions in the bottom panel of Fig.~\ref{fig:Tmaps_vs_absorption}.
     }
    \label{fig:plot_profile_eRObubble_T_NvsS}
\end{figure}
\begin{figure}
    \centering
    \includegraphics[width=0.49\textwidth]{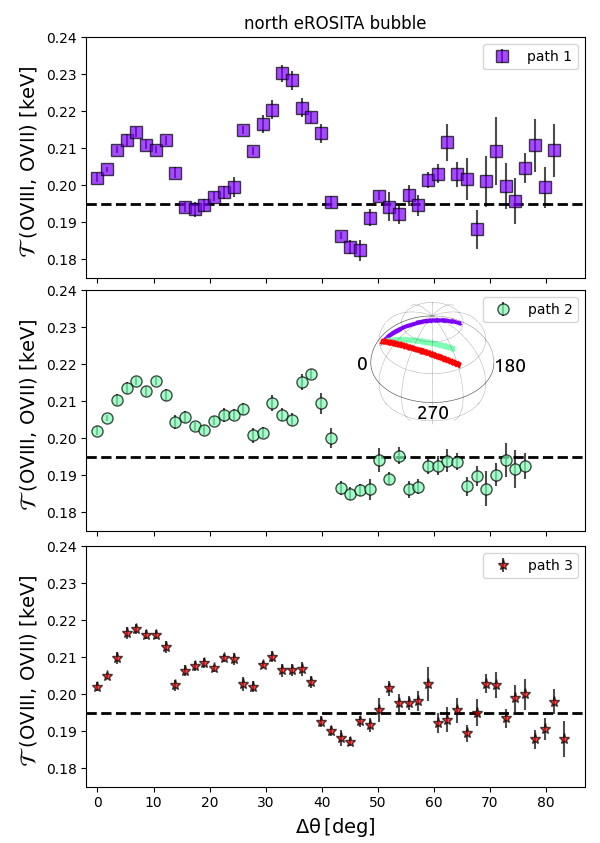}
    \caption{Pseudo-temperature profiles across the northern \erosita\ bubble. The paths are also shown as white transparent regions in the bottom panel of Fig.~\ref{fig:Tmaps_vs_absorption}.}
    \label{fig:plot_profile_eRObubbleN_paths123}
\end{figure}
To investigate in more detail the trend across the edge of the \erosita\ bubbles, we drew pseudo-temperature profiles, from $(l,b) = (350, \pm 15)$ to $(l,b) = (250, \pm 60)$. 
The two profiles shown in Fig.~\ref{fig:plot_profile_eRObubble_T_NvsS} are thus symmetric with respect to the Galactic plane, and their path in the sky is shown in the lower-left corner, and the bottom panel of Fig.~\ref{fig:Tmaps_vs_absorption}.
$\Delta \theta$ is the angle [deg] between a given point in the sky, along the profile, and the point of lower absolute latitude $|b|$.

The two profiles in Fig.~\ref{fig:plot_profile_eRObubble_T_NvsS} show different ratios between the regions inside (small $\Delta\theta$) and outside (large $\Delta\theta$, i.e. the CGM) the \erosita\ bubbles. In particular, the northern bubble (blue dots) shows a significant increase toward the inner regions, while a milder transition characterized the southern bubble (yellow squares). We stress once again that, as indicated by Fig. \ref{fig:spec}, the plasma inside the \erosita\ bubbles might have a contribution from a hotter component, which is not traced by the oxygen line ratio. Therefore, the mean temperature inside the bubbles is likely larger than the pseudo temperature shown in Fig. \ref{fig:plot_profile_eRObubble_T_NvsS}. 

Figure \ref{fig:plot_profile_eRObubble_T_NvsS} shows clear pseudo-temperature variations at the edges of the \erosita\ bubbles. 
Such transition happens at higher (absolute) latitudes in the northern bubble compared to the southern one. In other words, the X-ray emission from the bubbles is not symmetric with respect to the Galactic equator, with the northern bubble being brighter and more evident at higher latitudes. 
Complementary, while the foreground absorption in the northern hemisphere reaches $\rm N_H = 10^{21}\, cm^{-2}$ at $b\simeq 30-40$ deg, the southern hemisphere is less absorbed, with the same $N_H$ being limited to $b>-20$ deg. This allows us to extend the investigation on the pseudo-temperature in the southern bubble to regions closer to the Galactic disk or the Galactic center. 
In Fig.~\ref{fig:plot_profile_eRObubble_T_NvsS}, we partially hid the data points characterized by a column density $\rm N_H>10^{21}$~cm$^{-2}$, thus potentially biased by systematic on the absorption model.

The colder shell at about the boundary of the bubbles can be recognized by the steep drop at around $\Delta \theta \simeq 50-60$ deg from the path origin. 
In the south, the regions well inside the bubble are characterized by a higher average pseudo-temperature than in the north. 
A constant value is then observed well outside the boundaries eventually ($l\sim 240-210$ deg) matching the same level of the CGM regions (dashed lines).

The regions around the boundary of the bubbles show a systematic pseudo-temperature drop in both hemispheres when compared to the respective CGM values. 
The drop (i.e., the colder shell) is more evident in the north and is also present in the south. 
At the same time, a small but significant enhancement is observed $\Delta\theta=10-20$ deg before the drop in both hemispheres.
Strong deviations from homogeneity are found within the boundaries of the bubbles. 
While these inhomogeneities prevent us from characterizing the details of a putative pseudo-temperature jump, we find evidence for a higher pseudo-temperature and broad spatial variations of $\mathcal{T}$ inside the bubbles compared to the outer regions.

Given the brighter emission of the northern bubble, we further inspected the inhomogeneities through different profiles across it, as shown in the panels of Fig.~\ref{fig:plot_profile_eRObubble_T_NvsS} (the profiles are also charted as a white transparency in the bottom panel of Fig.~\ref{fig:Tmaps_vs_absorption}).
All the profiles have origin at $(l,b)=(350, 35)$. 
From these profiles, the features described above (e.g. the pseudo-temperature enhancement, the colder shell, the constant and scattered value at large $\Delta\theta$, and the large and coherent variations inside the bubbles) are all evident and highly significant. 
We note that along the profile encompassing the highest latitudes (Fig.~\ref{fig:plot_profile_eRObubbleN_paths123}, top panel), the pseudo-temperature shows a significant and increasing trend with respect to the average CGM value.
This confirms that the cold-warm shell noticed in the line ratio and pseudo-temperature maps is statistically significant. 

\section{Discussion} \label{sec:discussion}

Thanks to the \erosita\ energy resolution, we could derive the first half sky maps in narrow energy bands characteristic of some of the most prominent soft X-ray emission lines. 
These narrowband maps allowed us to detect several known features, reveal new structures, and use line ratio maps to derive a pseudo-temperature map of the CGM of the Milky Way. 

\subsection{On the pseudo-temperature distribution of the CGM}

The study of the soft X-ray emission lines (and their ratio), has allowed us to place constraints on the pseudo-temperature distribution of the CGM. 
We observe that on small angular scales, of few to $\sim20^\circ$, the CGM pseudo-temperature shows an overall normal distribution with fluctuations from the expected value of the order of a few percents, not accounted by our modeling. Conservatively, an overall upper limit to the intrinsic pseudo-temperature fluctuations of the order of $\Delta \mathcal{T}\leq 4$~\% can be considered. 
At first sight, this may appear in contrast with simulations of galaxies, predicting log-normal distributions of the CGM pseudo-temperatures and significant pseudo-temperature fluctuations from different regions in the sky. 
However, we note that we trace the pseudo-temperature distribution based only on the ratio of the \oe\ and \os\ lines, which are sensitive to the hotter component of the CGM. 
Therefore, we cannot exclude that the real pseudo-temperature distribution is actually skewed towards lower pseudo-temperature (e.g. traced by O {\tiny VI}), but we are insensitive to those CGM phases here. 

By comparing the CGM pseudo-temperature in the northern and southern hemispheres, we also observed significant pseudo-temperature variations on scales larger than $\sim20^\circ$, of the order of $\sim12$~\%. 
Significant pseudo-temperature variations are also observed when comparing the interior of the \erosita\ bubbles with the high Galactic latitude region away from the Galactic center. However, the presence of an additional hotter component prevents us from quantifying this pseudo-temperature jump in an unbiased way. 
Therefore, the study of the line ratio suggests that the CGM undergoes significant pseudo-temperatures variations of at least $\sim12$~\% on angular scales of several tens of degrees, while we observe that on smaller angular scales ($\sim2-20^\circ$) the pseudo-temperature is significantly smaller, and of the order of a few percent.

\subsection{An additional and hotter plasma component inside the \erosita\ bubbles}

In the previous Sect.~\ref{sec:bubble_profiles}, we detected clear pseudo-temperature variations at the edges of the \erosita\ bubbles.
The pseudo-temperature map, derived from the oxygen line ratio, indicates pseudo-temperatures of $\mathcal{T}\sim0.21-0.22$~keV and $\mathcal{T}\sim0.19-0.22$~keV for the region inside and outside the \erosita\ bubbles, respectively. 
This appears in agreement with some recent results based on the analysis of Suzaku spectra \citep{2023NatAs...7..799G}, which find a pseudo-temperature of the plasma inside the bubbles to be consistent with $\mathcal{T}\sim0.2$~keV, although with a significant overabundance of neon to oxygen of $\sim2$. We reiterate that the oxygen line ratio is not very sensitive to any component hotter than $T\sim0.3$~keV, as suggested by previous studies of the same Suzaku data, which did not require neon overabundance \citep{2018Galax...6...27K}. 
To discriminate between these different hypotheses, a full characterisation of the spectra from the inside and outside of the \erosita\ bubbles is necessary (e.g., Fig. \ref{fig:spec}).

\subsection{A shell surrounding the \erosita\ bubbles}

The investigation of the oxygen line ratio suggests the presence of a shell of colder material, with an inner hotter stripe, close to the edge of the \erosita\ bubbles, defined as the drop in surface brightness. 
The shell appears to wrap most of the northern bubble and a good fraction of the southern bubble closer to the disc. 
The origin of such a shell is still unclear. 
Colder phases as transition layers between an expanding bubble and the interstellar medium are typically observed in superbubbles, where a colder and high-density shocked interstellar gas is contained between the forward shock and the contact discontinuity separating it from the hotter and more rarefied shocked stellar wind \citep{1977ApJ...218..377W}. 
This colder and high-density shell forms because of radiative losses, which become relevant as the superbubble expands into the interstellar medium. 

Clearly, the example of superbubbles shows that an outflow interacting with the colder ambient gas can form a shell. 
However, the conditions experienced by the putative Galactic outflow, which might have shaped the \erosita\ bubbles, are probably closer to the ones in galactic outflows observed in star-forming and starburst galaxies, where the assumption of a homogeneous ambient medium creating a homogeneous shell in all direction is likely, not true. 
Indeed, the typical scale height of galactic outflows is significantly larger than their galactic discs. 
Therefore, the galactic disc is likely helping in impeding the expansion of the outflow along the disc and in collimating the outflow into a bipolar feature, which is expanding more towards directions where there is a significant drop in the density of the environment (the CGM). 
Because of the significant drop in the density of the environment (the CGM), outflows from star-forming galaxies are not expected to produce a colder cap, as observed in superbubbles.  
Instead, such denser and colder phases might be developed at the boundary between the outflow and the galactic disc and to trace the edges of the outflow up to several kiloparsecs above the galactic disc. 
Thus, within the framework of a galactic outflow, it is puzzling that the "colder shell" can be traced all the way to the top of the northern \erosita\ bubble. 

The observation of a shell reaching very high Galactic latitudes might be related to either entrainment and mixing between the hot and colder phases of the outflow or to projection effects. 
Indeed, the geometry of the outflow may be different from the one assumed in \cite{Predehl2020Natur} because the \erosita\ bubbles are so extended in the sky that the upper portion of this layer, which appears at Galactic latitudes of $b\sim60^\circ$ or more, may be associated with the near side of the Galactic outflow, which is so extended to fill the sky above us. 
We note that in such a scenario, it is somehow natural to expect that such a "colder shell" will trace the edges of the \erosita\ bubbles closer to the Galactic disc. 
Such a "colder shell" might be present all along the envelope around the \erosita\ bubble and closer to the Galactic disc, but to appear clearer at its edges because of projection effects.
If so, then no cap would be present at the top of the Galactic outflow and the "colder shell" observed in our maps would be due to the projection of the near side of the outflow, just a few kiloparsecs above the galactic disc. 

This prompts further consideration of the connections between \erosita\ bubbles and other known large-scale structures such as the North Pole Spur (NPS). This extended X-ray arc, previously detected by \rosat\ \citep{Snowden1997ApJ} emerges from the eastern galactic plane and gradually fades as it extends out from the disc. The NPS can be considered as part of the northern \erosita\ bubble. A bright and extended radio structure known as Loop-I has been associated to the NPS \citep{Berkhuijsen1971AA, Haslam1974AAS,Sofue1979AAS, Vidal2015MNRAS, Ade2016AA}. However, it is unclear whether the radio structure Loop-I and the northern eROSITA bubble share a common origin.
The high radio intensity from the low-latitude ISM regions complicates the determination of the origin of Loop I. 
There is still an open debate on the origin of the Loop-I/NPS feature. While some studies propose a Galactic Center (GC) origin \citep{Sofue1977AA, Sofue2000ApJ, Bland-Hawthorn2003ApJ, Kataoka2013ApJ}, others propose a local origin \citep{Egger1995AA, Dickinson2018Galax}. Currently, both scenarios are still consistent with observations. While the appearent lack of a corresponding southern feature for NPS/Loop-I posed a challenge for GC origin, the reassessment of the South Pole Spur (SPS) in planck map \citep{Ade2016AA} seemed to reduce the discrepancy with GC models. We note here that this radio-detected SPS aligns well with southern \erosita\ bubbles in the projection as viewed from Earth (see a review of \citet{Sarkar2024AARv}).

Of course, more work is needed to clarify the three-dimensional geometry of the \erosita\ bubbles and the origin of the features observed here. 
What seems clear from the data is the pseudo-temperature variation across the edges of the bubbles, indicating the presence of a shell. 

\section{Conclusions} \label{sec:sum}

We investigated the \erosita\ half sky maps within narrow energy bands characteristics of the brightest soft X-ray line transitions. 

\begin{itemize}
\item{} We subtracted the non-CGM components from soft X-ray emission lines (\os\ and \oe) and corrected the maps for the effects of absorption to determine the emission from the hot plasma associated with the Galactic CGM; 
\item{} Towards regions with little effects due to absorption, we derive maps of the pseudo-temperature of the CGM plasma from the \oe/\os\ line ratio, which is a good tracer of the pseudo-temperature of the warm-hot CGM unperturbed by the Galactic outflow;
\item{} We constrain the CGM temperature fluctuations for the first time on relatively small ($\sim2-20^\circ$) and large scale. Indeed, the CGM pseudo-temperature shows fluctuations not accounted by our background and foreground modeling, of the order of $\sim 3\%$, and with an upper limit of $\Delta \mathcal{T}/\mathcal{T}\le 4$~\% on angular scales between $\sim2$ and $\sim20^\circ$. Instead, the CGM temperature shows variations as large as $\sim12$~\% when comparing the northern and southern hemispheres, therefore indicating that indeed pseudo-temperature fluctuations are present on large angular scales (i.e., $\gg20^\circ$); 
\item{} Towards the northern and southern \erosita\ bubbles we detect significant pseudo-temperature fluctuations of order $2-3\%$ of the average values, indicating significant temperature fluctuations towards the interior of the bubbles. 
\item{} The line ratio map shows significant variations at the edges of the \erosita\ bubbles, suggesting the presence of a "shell" of lower values at the edge of the outflow. 
In particular, when the line ratio map is translated into pseudo-temperature map, we can see that the interior of the \erosita\ bubbles appears hotter than the surrounding CGM. 
Approaching the edge from the interior of the bubble, we observe a rise of the pseudo-temperature, followed by a significant drop, to then become again hotter, with a pseudo-temperature consistent with the surrounding CGM. 
This suggests the presence of a shell of cooler material present at the edge of the bubbles or of a separate process (not modeled in this work) affecting the different metal ionization states.
\end{itemize}

\begin{acknowledgements}
    This work is based on data from eROSITA, the soft X-ray instrument aboard SRG, a joint Russian-German science mission supported by the Russian Space Agency (Roskosmos), in the interests of the Russian Academy of Sciences represented by its Space Research Institute (IKI), and the Deutsches Zentrum für Luft- und Raumfahrt (DLR). The SRG spacecraft was built by Lavochkin Association (NPOL) and its subcontractors, and is operated by NPOL with support from the Max Planck Institute for Extraterrestrial Physics (MPE).
    The development and construction of the eROSITA X-ray instrument was led by MPE, with contributions from the Dr. Karl Remeis Observatory Bamberg \& ECAP (FAU Erlangen-Nuernberg), the University of Hamburg Observatory, the Leibniz Institute for Astrophysics Potsdam (AIP), and the Institute for Astronomy and Astrophysics of the University of Tübingen, with the support of DLR and the Max Planck Society. The Argelander Institute for Astronomy of the University of Bonn and the Ludwig Maximilians Universität Munich also participated in the science preparation for eROSITA.
    The eROSITA data shown here were processed using the eSASS/NRTA software system developed by the German eROSITA consortium.

    We acknowledge financial support from the European Research Council (ERC) under the European Union’s Horizon 2020 research and innovation program Hot- Milk (grant agreement No 865637). GP acknowledges support from Bando per il Finanziamento della Ricerca Fondamentale 2022 dell’Istituto Nazionale di Astrofisica (INAF): GO Large program and from the Framework per l’Attrazione e il Rafforzamento delle Eccellenze (FARE) per la ricerca in Italia (R20L5S39T9). MF and MY acknowledge support from the Deutsche Forschungsgemeinschaft through the grant FR 1691/2-1.
\end{acknowledgements}

\bibliography{aa}

\appendix

\section{Foreground absorption} \label{sec:absorption}

\begin{figure}
    \centering
    \includegraphics[width=0.5\textwidth]{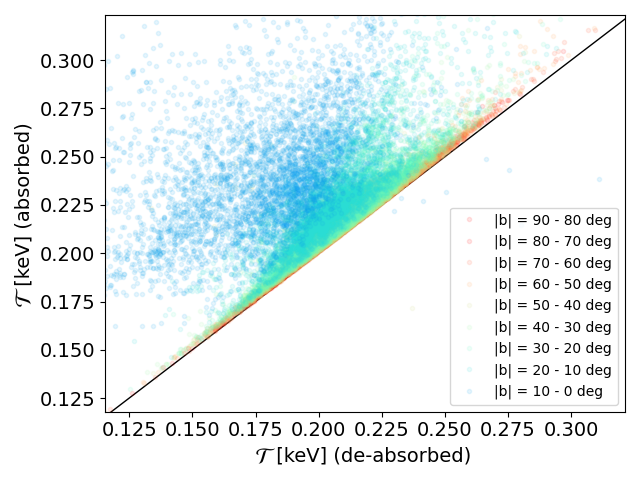}
    \caption{Pseudo-temperature values derived by the absorbed (i.e. upper panel) versus the de-absorbed (i.e. Fig.~\ref{fig:RatO}) lines. The pixel values are colored accoring to their absolute Galactic latitude, by ranges of 10 deg, as labelled. All $|b|>30$ deg values deviate from the 1:1 relation (black solid line) to less than 5\% of their value.}
    \label{fig:Tmaps_vs_absorption_plot}
\end{figure}
The flux of an emission line is contributed by different sources, as discussed above. Some of these sources, such as the CXB and the CGM (i.e. the object of our study) are also absorbed by foreground cold(er) layers of material. In addition, the absorption coefficient depends on the energy considered \citep{1992ApJ...400..699B}. To correctly interpret the line ratio between different emission lines it is thus necessary to account for the fraction of absorbed ration and add it back to the total emitted line flux. In order to do so, a model for the absorption column density (and metal content) has to be assumed. The final line ratio thus depends on the absorption model, and potentially introduces systematics in the results. 
The absorption model adopted in this work is the same as the one described in \cite{2023arXiv231010715L}, we refer the reader to that work for further details on the model.
In Fig.~\ref{fig:Tmaps_vs_absorption} we plot a pseudo-temperature map (upper panel) derived from a line ratio where we did not added back the absorbed emission, and thus does not depend on the absorption model. By comparing it with Fig.~\ref{fig:RatO} (we plot one versus the other in the lower panel of Fig.~\ref{fig:Tmaps_vs_absorption}) we show that for all absolute Galactic latitudes $|b|\geq 20$ deg, the deviation introduced by the absorption model are negligible. This is mostly due to the relatively small column densities of absorbing material found at these latitudes. This also means that any systematic deviation introduced by the absorption model will be smaller than the deviation from unity seen in the lower panel Fig.~\ref{fig:Tmaps_vs_absorption} at a given latitude. Given that our representative samples in both the northern and southern hemispheres have been drawn at $|b|>30$ deg for all pixels, we constrain any systematic deviation introduced by the absorption model in our samples to $\Delta \mathcal{T}/\langle \mathcal{T} \rangle <5\%$.

\end{document}